\documentclass[journal]{IEEEtran}

\ifCLASSINFOpdf
\usepackage[pdftex]{graphicx}
\DeclareGraphicsExtensions{.pdf,.jpeg,.png}
\else
\usepackage[dvips]{graphicx}
\fi
\usepackage{subfigure}
\usepackage{epstopdf}
\usepackage[cmex10]{amsmath}
\usepackage{amssymb}
\usepackage{wasysym}
\usepackage{algorithm}
\usepackage{algorithmic}
\usepackage{array}
\usepackage{cite}
\usepackage{color}
\usepackage{url}
\usepackage{bm}
\usepackage{enumerate}
\usepackage{epsfig,latexsym}%amsmath
\usepackage{cite}
\usepackage{hyperref}
\begin{document}
\title{Multiuser Beamforming for Partially-Connected Millimeter Wave Massive MIMO}

%\author{Kangjian~Chen~\authorrefmark{1}\authorrefmark{2}, Chenhao~Qi~\authorrefmark{1}\authorrefmark{2} and Octavia A. Dobre~\authorrefmark{3}\\
%	\authorrefmark{1}National Mobile Communications Research Laboratory\\
%	\authorrefmark{2}School of Information Science and Engineering, Southeast University, Nanjing, China  \\
%	\authorrefmark{3}Faculty of Engineering and Applied Science, Memorial University, Canada\\
%	Email:~\{kjchen,~qch\}@seu.edu.cn,~odobre@mun.ca
%}

\author{Chenhao~Qi,~\IEEEmembership{Senior~Member,~IEEE}, Jinlin~Hu, Yang Du,
	Arumugam Nallanathan,~\IEEEmembership{Fellow,~IEEE}
	\thanks{Copyright (c) 2015 IEEE. Personal use of this material is permitted. However, permission to use this material for any other purposes must be obtained from the IEEE by sending a request to pubs-permissions@ieee.org.}
	\thanks{This work was supported in part by the National Natural Science Foundation of China under Grants 62071116 and U22B2007. (\textit{Corresponding author: Chenhao~Qi})}
	\thanks{Chenhao~Qi and Jinlin~Hu are with School of Information Science and Engineering, Southeast University, Nanjing 210096, China (Email: \{jlhu, qch\}@seu.edu.cn).}
	\thanks{Yang Du is with Huawei Technologies Company Ltd. (Chengdu), Chengdu, China (Email: duyang22@huawei.com).}
	\thanks{Arumugam Nallanathan is with School of Electronic Engineering and Computer Science, Queen Mary University of London, London, UK (Email: nallanathan@ieee.org).}
}

\markboth{Accepted by IEEE Transactions on Vehicular Technology}
{}

\maketitle

\begin{abstract}
Multiuser beamforming is considered for partially-connected millimeter wave massive MIMO systems. Based on perfect channel state information (CSI), a low-complexity hybrid beamforming scheme that decouples the analog beamformer and the digital beamformer is proposed to maximize the sum-rate. The analog beamformer design is modeled as a phase alignment problem to harvest the array gain. Given the analog beamformer, the digital beamformer is designed by solving a weighted minimum mean squared error problem. Then based on imperfect CSI, an analog-only beamformer design scheme is proposed, where the design problem aims at maximizing the desired signal power on the current user and minimizing the power on the other users to mitigate the multiuser interference. The original problem is then transformed into a series of independent beam nulling subproblems, where an efficient iterative algorithm using the majorization–minimization framework is proposed to solve the subproblems. Simulation results show that, under perfect CSI, the proposed scheme achieves almost the same sum-rate performance as the existing schemes but with lower computational complexity; and under imperfect CSI, the proposed analog-only beamforming design scheme can effectively mitigate the multiuser interference.

\end{abstract}
\begin{IEEEkeywords}
Analog beamforming, hybrid beamforming, millimeter wave (mmWave) communications, partially-connected structure.
\end{IEEEkeywords}

\section{Introduction}
Although millimeter wave (mmWave) communications have been widely recognized as an important candidate for future wireless systems owing to the rich spectral resource in mmWave frequency band, they have not yet been commercially deployed for cellular wireless networks~\cite{SCICH2021CSIQi}. The barriers include the high cost in hardware and the large energy consumption especially for mobile terminals~\cite{TSTSP2016Altopt}. Though the fully-connected structure is more intensively studied by the academia and has been shown to achieve better performance than the partially-connected structure, the latter has much lower hardware complexity, e.g., much smaller number of phase shifters, and is more preferable by the industry than the former~\cite{PartiallyConnected}.

% The short wavelength of mmWave signal enables a large antenna array to be packed into a small area, which facilitates massive MIMO transmission to compensate for the path loss induced by high frequency. However, due to the large power consumption and hardware cost, it is impractical to allocate a dedicated radio frequency (RF) chain for each antenna in mmWave massive MIMO systems. Consequently, a hybrid architecture where a small number of RF chains are connected to a large number of antennas through phase shifter networks is usually adopted. According to the mapping from RF chains to antennas, the hybrid structure can generally be divided into the fully-connected and ~\cite{CM2017Survey}. Although the fully-connected architecture has the potential to achieve higher beamforming gain, it requires more complex circuitry and relatively high power than partially-connected architectures.

For multiuser mmWave massive MIMO, hybrid beamforming is generally adopted, where the analog beamformer implemented by the phase shifters achieves directional signal transmission and the digital beamformer mitigates the interference among different users. In~\cite{alkhateeb2015limited}, aiming at maximizing the sum-rate, a two-stage limited feedback multiuser hybrid beamforming (TSH) scheme is proposed, where the analog beamformer is first designed by beam sweeping and then the base station (BS) trains the effective channels to perform zero-forcing processing with the designed digital beamformer. To further increase the sum-rate, the analog beamformer is designed by performing singular value decomposition (SVD) of the high-dimensional air-interface channel matrix, 
while the digital beamformer is designed through block diagonalization~\cite{wu2018hybrid}. In~\cite{SIC}, the analog beamformer is designed the same as~\cite{wu2018hybrid} while the digital beamformer is designed based on successive optimization; and in particular, the digital beamformer of the current user is required to lie in the null space of the effective channels of the remaining users so that the multiuser interference can be mitigated. Note that both~\cite{wu2018hybrid} and \cite{SIC} assume that the ideal channel state information (CSI) is available to the hybrid beamformer design.

In this paper, we consider the multiuser beamforming for partially-connected mmWave massive MIMO based on perfect CSI and imperfect CSI, respectively. Our contributions can be summarized as:
\begin{itemize}
	\item Based on perfect CSI, we propose a low-complexity hybrid beamforming scheme that decouples the analog beamformer and the digital beamformer to maximize the sum-rate. The analog beamformer design is modeled as a phase alignment problem to harvest the array gain. Given the analog beamformer, the digital beamformer is designed by solving a weighted minimum mean squared error problem.
	\item Since perfect CSI is not available in practice, then based on imperfect CSI, an analog-only beamformer design scheme is proposed, where the design problem aims at maximizing the desired signal power on the current user and minimizing the power on the other users to mitigate the multiuser interference. The original problem is then transformed into a series of independent beam nulling subproblems, where an efficient iterative algorithm using the majorization–minimization framework is proposed to solve the subproblems. Different from the existing work, our scheme completely omits the digital beamformer design, so that the pilot overhead in estimating the effective channels can be completely removed.
\end{itemize}

The rest of the paper is organized as follows. Section~\ref{sec.System Model} introduces the system model of partially-connected
mmWave massive MIMO. In Section~\ref{sec.HybridPrecoding}, we investigate the hybrid beamforming based on perfect CSI. Since the perfect CSI is unavailable in practice, we then investigate the analog-only beamforming design based on imperfect CSI in Section~\ref{AnalogOnly}. The simulation results are presented in Section~\ref{sec.SimulationResults}, and the paper
is concluded in Section~\ref{sec.Conclusion}.

\textit{Notations}: Symbols for vectors (lower case) and matrices (upper case) are in boldface. For a vector $ \boldsymbol{a} $, $ \left[\boldsymbol{a} \right]_m  $denotes its $ m $th entry. For a matrix $ \boldsymbol{A} $, $ \left[ \boldsymbol{A}\right]_{m,:} $, $ \left[\boldsymbol{A}\right]_{:,n} $, $ \left[ \boldsymbol{A}\right]_{m,n} $, $ \boldsymbol{A}^T $, $ \boldsymbol{A}^{-1} $, $ \boldsymbol{A}^H $ and $\big\| \boldsymbol{A} \big\|_{F} $ denote the $ m $th row, $ n $th column, entry on the $ m $th row and $ n $th column, transpose, inverse, conjugate transpose (Hermitian), and Frobenius norm, respectively. $ \boldsymbol{I}_K $ and $\mathcal{CN}(0,\sigma^2)$ denote an identity matrix of size $ K $ and the complex Gaussian distribution with zero mean and the variance being $ \sigma^2 $, respectively. The symbols $ \angle\left\lbrace \cdot \right\rbrace  $, $ {\rm Re} \left\lbrace \cdot \right\rbrace  $ and $|\cdot|$ denote the angle, the real part and the absolute value of a complex-valued number, respectively. The symbols $ \lfloor \cdot \rfloor $, $ {\rm blkdiag}\left\lbrace \cdot\right\rbrace$, $\mathbb{E}\{\cdot\}$ and $ \mathbb{C} $ denote the floor operation, block diagonalization operation, expectation and set of complex-valued numbers, respectively.

%%%%%%%%%%%%%%%%%%%%%%%%%%%%%%%%%%%%%%%%%%%%%%%%%%%%%%%%%%%%%%%%%%%%%%%%%%%%%%%%%%%%%%%%%%%%%%%%%%%%%%%%%%%%%%%%%%%%%%%%%%%%%%%%%%%%%%%%%%%%%%%%%%%%%%%%%%%%%%%%%%

\section{System Model}\label{sec.System Model}

%% 系统模型图
%\begin{figure}[!t]
%	\centering
%	\includegraphics[width=90mm]{SYSTEM_NEW.jpg}
%	\caption{A BS with the partially-connected hybrid beamforming architecture communicating with $ K $ users.}
%	\label{fig:system}
%\end{figure}

%It is commonly assumed that the number of simultaneously served users is the same as that of BS RF chains to save power consumption. For simplicity, we also assume that the BS will use $ K $ out of the $ N_{\rm RF} $ available RF chains to serve the $ K $ users.
We consider the downlink transmission of a mmWave massive MIMO, where the BS equipped with $N_{\rm BS}$ uniform linear array (ULA) antennas and $N_{\rm RF}$ radio frequency (RF) chains ($ N_{\rm BS} \gg N_{\rm RF} \geqslant 1 $) is assumed to serve $K$ single antenna users. The number of users simultaneously served by the BS is constrained by the number of RF chains, i.e., $ K\leqslant N_{\rm RF} $. In this paper, we assume $ K=N_{\rm RF} $ for simplicity.\footnote{In practice, if the number of mobile devices is greater than the number of RF chains, i.e., $K>N_{\rm RF}$, we can employ some user scheduling algorithms, which select $N_{\rm RF}$ users from $K$ candidate users to serve simultaneously at a time, and select another $N_{\rm RF}$ users to serve at another time by time-division multiple access (TDMA).}
The hybrid beamformer at the BS includes a digital beamformer at baseband (BB) and an analog beamformer in the RF domain, denoted as $ \boldsymbol{F}_{\rm BB}\in \mathbb{C}^{K \times K}$ and $ \boldsymbol{F}_{\rm RF} \in \mathbb{C}^{N_{\rm BS} \times K} $, respectively. Moreover, we normalize the power gain of the hybrid beamformer by setting $ \| \boldsymbol{F}_{\rm RF} \boldsymbol{F}_{\rm BB} \|^2_F = K $ to ensure that the hybrid beamformer does not provide any power gain. The received signal by the $ k $th user can be expressed as 
\begin{equation}\label{system model}
	y_k = \boldsymbol{h}^T_k\boldsymbol{F}_{\rm RF}\boldsymbol{F}_{\rm BB}\boldsymbol{s} + \eta_k,
\end{equation}
where $\boldsymbol{s} \triangleq \left[s_1, s_2, \ldots, s_K\right]^T\in\mathbb{C}^{K}$ is a data symbol vector subject to the constraint of total transmit power $ P $, i.e., $ \mathbb{E}\left\{ \boldsymbol{s}\boldsymbol{s}^H \right\} = \frac{P}{K}\boldsymbol{I}_K $, and $ \eta_k \sim \mathcal{CN} \left( 0, \sigma^2 \right) $ denotes the additive white Gaussian noise. $ \boldsymbol{h}_k \in \mathbb{C}^{N_{\rm BS}} $ denotes the channel vector between the BS and the $k$th user. 

According to the widely-used Saleh-Valenzuela model~\cite{alkhateeb2015limited}, which takes into account the limited scattering characteristics in mmWave channels and assumes a geometric channel model with $L_k$ independent propagation paths, the channel vector $\boldsymbol{h}_k$ in \eqref{system model} can be defined as 
\begin{equation}\label{channelmodel}
    \boldsymbol{h}_k=\sqrt{\frac{N_{\rm BS}}{L_k}} \sum_{l=1}^{L_k}\alpha_{k,l}\boldsymbol{a}(\theta_{k,l})  
\end{equation}
where $ \alpha_{k,l} $ and $ \theta_{k,l} \triangleq {\rm sin}(\vartheta_{k,l})$ denote the complex gain and the angle-of-departure (AoD) of the $l$th path, respectively. $ \vartheta_{k,l} \in \left( -\pi/2,\pi/2\right]$ denotes the physical angle for the AoD and $ L_k $ denotes the number of channel paths. Due to the significant path loss caused by mmWave frequency band, the number of transmit antennas is much higher than the number of resolvable paths, i.e., $ N_{\rm BS} \gg L_k$. The function $ \boldsymbol{a}\left( \cdot\right) $ denotes the array steering vector and can be expressed as 
$	\boldsymbol{a}(\theta)\triangleq{\big[1,e^{j\pi \theta},e^{j2\pi\theta},\ldots,e^{j(N_{\rm BS}-1)\pi \theta}\big]}^T /\sqrt{N_{\rm BS}} $
for ULAs with a half-wavelength antenna space.

Note that the BS employs partially-connected structure, where each RF chain is only connected with $N_{\rm s} = [N_{\rm BS}/ N_{\rm RF}]$ antennas. Then $ \boldsymbol{F}_{\rm RF} $ can be denoted as a block-diagonal matrix
$\boldsymbol{F}_{\rm RF} \triangleq {\rm blkdiag}\left\lbrace  \boldsymbol{f}_1, \boldsymbol{f}_2, \ldots, \boldsymbol{f}_K \right\rbrace$, where $ \boldsymbol{f}_k \in \mathbb{C}^{ N_{\rm s}}$ for $k=1,2,\ldots,K$ denotes the analog beamforming vector corresponding to the $ k $th antenna subarray. Since $ \boldsymbol{F}_{\rm RF} $ is implemented with phase shifters~\cite{LZH}, each non-zero entry of $ \boldsymbol{F}_{\rm RF} $ has a constant modulus constraint, i.e., 
\begin{equation}
	\big\lvert [\boldsymbol{f}_{k}]_{n} \big\rvert  = \frac{1}{\sqrt {N_{\rm s}}}, n=1,2,\ldots,N_{\rm s}, k=1,2,\ldots,K.	
%	1\leqslant n \leqslant N_{\rm s}, 1\leqslant k \leqslant K.
\end{equation}

%%%%%%%%%%%%%%%%%%%%%%%%%%%%%%%%%%%%%%%%%%%%%%%%%%%%%%%%%%%%%%%%%%%%%%%%%%%%%%%%%%%%%%%%%%%%%%%%%%%%%%%%%%%%%%%%%%%%%%%%%%%%%%%%%%%%%%%%%%%%%%%%%%%%%%%%%%%%%%%%%%

\section{Hybrid Beamforming Based on Perfect CSI}\label{sec.HybridPrecoding}
In this section, we investigate the hybrid beamformer design based on perfect CSI. The optimal hybrid beamformer at the BS is typically designed to maximize the sum-rate of all users by solving the following problem expressed as
\begin{subequations}\label{PerfectCSIProblem}
	\begin{align}
		& \max_{\boldsymbol{F}_{\rm RF},\boldsymbol{F}_{\rm BB}} {\sum_{k=1}^{K} R_k}\\
		& ~~~~{\rm s.t.} \ \ \ {{\big\|\boldsymbol{F}_{\rm RF}\boldsymbol{F}_{\rm BB}\big\|}^2_F = K},\\
		& \ \ \ \ \ \ \ \ \ \ \ \boldsymbol{F}_{\rm RF} = {\rm blkdiag}\left\lbrace \boldsymbol{f}_1, \boldsymbol{f}_2, \ldots, \boldsymbol{f}_K \right\rbrace, \\
		& \ \ \ \ \ \ \ \ \ \ \ \lvert \left[\boldsymbol{f}_{k}\right]_{n} \rvert  = \frac{1}{\sqrt {N_{\rm s}}}, n=1,2,\ldots,N_{\rm s}, k=1,2,\ldots,K,
		%1\leqslant n \leqslant N_{\rm s}, 1\leqslant k \leqslant K,
	\end{align}
\end{subequations}
where
\begin{equation}\label{Rk}
	R_{k} = \log_2\left( 1 + \frac{\frac{P}{K} \left| \boldsymbol{h}^{H}_{k}\boldsymbol{F}_{\rm RF} \left[ \boldsymbol{F}_{\rm BB} \right]_{:,k} \right|^2}
	{\frac{P}{K} \sum_{i\neq k}^{K} \left| \boldsymbol{h}^{H}_{k}\boldsymbol{F}_{\rm RF} \left[ \boldsymbol{F}_{\rm BB} \right]_{:,i} \right|^2 + \sigma^2} \right), 
\end{equation}
denotes the achievable rate of the $ k $th user.
%\begin{equation}\label{FRF}
%	\boldsymbol{F}_{\rm RF} = 
%	\begin{bmatrix}
%		\boldsymbol{f}_{1} & \boldsymbol{0}  & \cdots & \boldsymbol{0} \\
%		\boldsymbol{0}  & \boldsymbol{f}_{2} & \cdots & \boldsymbol{0}\\
%		\vdots & \vdots  & \ddots & \vdots\\
%		\boldsymbol{0}& \boldsymbol{0}  & \cdots &\boldsymbol{f}_{K}
%	\end{bmatrix},
%\end{equation}

Note that the design of the analog beamformer and digital beamformer is coupled in \eqref{PerfectCSIProblem}, which is difficult to handle. Now we consider a hybrid beamforming scheme that can decouple $ \boldsymbol{F}_{\rm RF} $ and $ \boldsymbol{F}_{\rm BB} $. Specifically, the analog beamformer is first designed, which is used to harvest the array gain provided by the antennas in mmWave massive MIMO. Then fixing the analog beamformer, the digital beamformer is designed to maximize the sum-rate of effective channels.

\subsubsection{Analog Beamformer Design}\label{subsubsec.PerfectCSIAnalog}
Since the perfect CSI is assumed to be available to the BS, the harvest of array gain can be achieved by aligning the phases of the analog beamformer to the phases of the corresponding channel entries. In particular, the analog beamformer can be optimized by 
\begin{equation}\label{FRFSolve}
	\begin{aligned}
		&\left[ \boldsymbol{F}_{\rm RF} \right]_{t, k} = \frac{1}{\sqrt{N_{\rm s}} }e^{j\theta_{t,k}},\\
		&k=1,2,\ldots, K,~t=(k-1)N_{\rm s}+1,\ldots,kN_{\rm s},
	\end{aligned}
\end{equation}
where $ \theta_{t,k} $ is the phase of the $ t $th entry of $ \boldsymbol{h}_k $.

\subsubsection{Digital Beamformer Design}\label{subsubsec.PerfectCSIDigital}
Given the optimized $ \boldsymbol{F}_{\rm RF} $, the effective channel $ \boldsymbol{\widetilde{h}}_k \in \mathbb{C}^{K}$ for the $ k $th user can be expressed as~\cite{SXY}
\begin{equation}\label{hkeffective}
	\boldsymbol{\widetilde{h}}_k = \boldsymbol{F}^{H}_{\rm RF}\boldsymbol{h}_k.
\end{equation}
Then we can determine $ \boldsymbol{F}_{\rm BB} $ by rewriting \eqref{PerfectCSIProblem} as
\begin{subequations}\label{PerfectCSIFBB}
	\begin{align}
		& \max_{\boldsymbol{F}_{\rm BB}} {\sum_{k=1}^{K} 
			\log_2\left( 1 + \frac{\frac{P}{K} \left| \widetilde{\boldsymbol{h}}^{H}_{k}\left[ \boldsymbol{F}_{\rm BB} \right]_{:,k} \right|^2}
			{\frac{P}{K} \sum_{i\neq k}^{K} \left| \widetilde{\boldsymbol{h}}^{H}_{k} \left[ \boldsymbol{F}_{\rm BB} \right]_{:,i} \right|^2 + \sigma^2} \right) }\\
		& ~{\rm s.t.} \ {{\big\|\boldsymbol{F}_{\rm BB}\big\|}^2_F = K}.
	\end{align}
\end{subequations}
% ~\cite[Thm. 1]{shi2011iterativelyprovesource}
By exploiting the relationship between sum-rate maximization and minimum mean squared error (MMSE)~\cite{shi2011iterativelyprovesource}, we can equivalently convert \eqref{PerfectCSIFBB} into a weighted MMSE (WMMSE) problem as
\begin{subequations}\label{PerfectCSIFBBEq}
	\begin{align}
		& \min_{\boldsymbol{F}_{\rm BB},\left\lbrace w_k, u_k\right\rbrace^K_{k = 1}} 
		\sum_{k=1}^{K} \{w_{k}e_{k} - \log_2{w_k}\}\\
		& ~~~~~~~~{\rm s.t.} \ \ \ \ \ \ {{\big\|\boldsymbol{F}_{\rm BB}\big\|}^2_F = K}\label{FBBconstraint},
	\end{align}
\end{subequations}
where $ w_k $ is a positive weight variable, $ u_k $ is an auxiliary variable, and $ e_k $ is the mean squared error defined as
\begin{equation}\label{ek}
	\begin{split}
		e_k =& \frac{P}{K}\left| 1-u_k\widetilde{\boldsymbol{h}}^{H}_{k}\left[ \boldsymbol{F}_{\rm BB} \right]_{:,k} \right|^2\\
		&+ \frac{P}{K}\sum_{i\neq k}^{K} \left| u_k\widetilde{\boldsymbol{h}}^{H}_{k}\left[ \boldsymbol{F}_{\rm BB} \right]_{:,i} \right|^2 
		+\sigma^2\left|u_k\right|^2.
	\end{split}
\end{equation}

Since $ w_k $, $ u_k $ and $ \boldsymbol{F}_{\rm BB} $ are coupled in \eqref{PerfectCSIFBBEq}, it can be solved by exploiting alternating optimization based on the method proposed in~\cite{shi2011iterativelyprovesource}, where $ u_k $, $ w_k $ and $ \left[ \boldsymbol{F}_{\rm BB} \right]_{:,k} $  can be updated by 
\begin{subequations}\label{AltOptIter}
	\begin{align}
		u_k &= \frac{\widetilde{\boldsymbol{h}}^{H}_{k}\left[ \boldsymbol{F}_{\rm BB} \right]_{:,k}}
		{\sum\nolimits_{i\neq k}^K {| \widetilde{\boldsymbol{h}}^{H}_{k}\left[ \boldsymbol{F}_{\rm BB} \right]_{:,i} |^2} +\sigma^2},\label{uk}\\
		w_k &= \frac{1}{e_k},\label{wk}\\
		\left[ \boldsymbol{F}_{\rm BB} \right]_{:,k} &= w_{k}u_{k}
		\left(\sum_{i=1}^K{w_{i}|u_{i}|^2\widetilde{\boldsymbol{h}}_{i}\widetilde{\boldsymbol{h}}^{H}_{i}}+\mu\boldsymbol{I}_{K}\right)^{-1} 
		\widetilde{\boldsymbol{h}}_{k},\label{FBBk}
	\end{align}
\end{subequations}
in each iteration. In fact, $ \mu $ in~\eqref{FBBk} is known as the Lagrangian multiplier associated with~\eqref{FBBconstraint}.

The overall hybrid beamforming scheme based on perfect CSI for solving~\eqref{PerfectCSIProblem} is summarized as  \textbf {Algorithm~\ref{alg1}}, which is named as the phase-alignment WMMSE (P-WMMSE) scheme. In \textbf {Algorithm \ref{alg1}}, we normalize $ \boldsymbol{F}_{\rm BB} $ by
\begin{equation}\label{FBBnormalize}
	\left[\boldsymbol{F}_{\rm BB}\right]_{:,k} = \frac{\left[ \boldsymbol{F}_{\rm BB} \right]_{:,k}}{{\left\|\left[ \boldsymbol{F}_{\rm BB} \right]_{:,k}\right\|}_2} , k=1,2,\ldots,K,
\end{equation}
in step~\ref{FBBn1} to satisfy~\eqref{FBBconstraint}. The \textit{stop condition} in step~\ref{FBBed} can be set as either the number of iterations exceeding a predefined value or the relative difference between the solutions obtained by two consecutive iterations becoming smaller than a specified threshold.

The computational complexity of the P-WMMSE scheme mainly includes two parts. The first part is the phase extraction in solving $ \boldsymbol{F}_{\rm RF} $, and the second part is the alternating optimization in solving $ \boldsymbol{F}_{\rm BB} $. Since the number of phases to be computed in the first part is $ N_{\rm BS} $, the computational complexity of phase extraction is $ \mathcal{O}\left(N_{\rm BS}\right) $. We denote the predefined number of iterations for the alternating optimization as $ N_{\rm max} $. According to~\eqref{AltOptIter}, the computational complexity for computing $ u_k $, $ w_k $ and $ \left[ \boldsymbol{F}_{\rm BB} \right]_{:,k} $ in together is $ \mathcal{O}\left( K^3\right) $ during each iteration. Then the total computational complexity for the P-WMMSE scheme is $ \mathcal{O}\left(N_{\rm BS} + N_{\rm max}K^4\right) $. By contrast, the computational complexity of the HySBD is $ \mathcal{O}\left( N_{\rm BS}^3 + K^4\right)$~\cite{wu2018hybrid}. In generally setting with $N_{\rm BS} \geq 64$, $N_{\rm max} \leq 20$ and $K \leq 8$, the computational complexity of P-WMMSE is much lower than that of 
HySBD.

%\begin{equation}
%	\mathcal{O}\left(N_{\rm BS} + N_{\rm max}K^4\right).
%\end{equation}
%By contrast, the complexity of the HySBD~\cite{wu2018hybrid} is 
%\begin{equation}
%	\mathcal{O}\left( N_{\rm BS}K^2 + K^4\right).
%\end{equation}
%Thus, by setting an suitable value of $ N_{\rm max} $, the proposed P-WMMSE hybrid beamforming scheme can enjoy lower complexity.

\begin{algorithm}[!t]
	\caption{P-WMMSE Hybrid Beamforming Scheme Based on Perfect CSI}
	\label{alg1}
	\begin{algorithmic}[1]
		\STATE \textbf{Input:} $ \left\lbrace \boldsymbol{h}_k \right\rbrace ^K_{k=1} $, $ \sigma^2 $, 
		$ N_{\rm BS} $, $ N_{\rm s} $.
		% FRF
		\FOR{$k=1:K$}\label{FRFst}
			\STATE Obtain $\left[ \boldsymbol{F}_{\rm RF} \right]_{t, k}$ via \eqref{FRFSolve}, $t =(k-1)N_{\rm s}+1,\ldots, kN_{\rm s} $.
		\ENDFOR\label{FRFed}
		
		\STATE Obtain $\widetilde{\boldsymbol{h}}_{k}$ via (\ref{hkeffective}), $ \forall k $.
		\STATE Initialize $\boldsymbol{F}_{\rm BB} = [\widetilde{\boldsymbol{h}}_{1}, \widetilde{\boldsymbol{h}}_{2}, \ldots, \widetilde{\boldsymbol{h}}_{K}]$.
		\STATE Normalize $\boldsymbol{F}_{\rm BB}$ via (\ref{FBBnormalize}).\label{FBBn1}
		% FBB
		\REPEAT\label{FBBst}
		\STATE Obtain $ u_k $, $ e_k $, $ w_k $ and $ \left[ \boldsymbol{F}_{\rm BB}\right]_{:,k} $ via (\ref{uk}), (\ref{ek}), (\ref{wk}) and (\ref{FBBk}), respectively, $ \forall k $.
		\UNTIL \textit{stop condition}  is satisfied.\label{FBBed}
		\STATE \textbf{Output:} $\boldsymbol{F}_{\rm BB}$, $\boldsymbol{F}_{\rm RF}$.
	\end{algorithmic}
\end{algorithm}
%%%%%%%%%%%%%%%%%%%%%%%%%%%%%%%%%%%%%%%%%%%%%%%%%%%%%%%%%%%%%%%%%%%%%%%%%%%%%%%%%%%%%%%%%%%%%%%%%%%%%%%%%%%%%%%%%%%%%%%%%%%%%%%%%%%%%%%%%%%%%%%%%%%%%%%%%%%%%%%%%%%

\section{Analog-only Beamforming Based on Imperfect CSI}\label{AnalogOnly}
The P-WMMSE scheme is based on perfect CSI. Since the perfect CSI according to \eqref{channelmodel} includes the number of channel paths, the complex gain and AoD of each channel path, acquiring perfect CSI in practice is challenging. Typically we use the beam sweeping to obtain the best codeword for each user, where each codeword covers an angle range of the AoD. However, the beam sweeping cannot directly acquire an estimate of the CSI. The number of channel paths and the complex gain are completely unknown after beam sweeping. To acquire the CSI, additional overhead  for channel parameter estimation is needed besides of the beam sweeping. Therefore, the P-WMMSE scheme cannot be performed only based on the beam sweeping. 

In this section, we investigate the analog-only beamformer design based on the beam sweeping. Note that different from the existing work on hybrid beamformer design, our work completely omits the digital beamformer design, so that the pilot overhead in estimating the effective channels can be completely removed.

% corresponding to each user, which is what imperfect CSI refers to and the premise of analog-only beamformer design in this paper. The motivation for this part of the work comes from the following two aspects: (1) It is challenging to obtain perfect CSI in practical applications, and even if channel estimation techniques can obtain relatively accurate CSI, the required pilot overhead is relatively large; (2) Designing digital beamformer with the estimated low-dimensional effective channel requires additional pilot overhead besides beam sweeping or beam training. Therefore, from the perspective of reducing the difficulty of CSI acquisition and pilot overhead, we investigate the analog-only beamformer design with imperfect CSI.

%, which refers to the codewords corresponding to each user obtained by beam sweeping rather than the perfect high-dimensional channel $ \left\lbrace \boldsymbol{h}_k \right\rbrace ^K_{k=1} $.

We still adopt the same system model, where we set $ \boldsymbol{F}_{\rm BB} = \boldsymbol{I}_K $ to indicate that the digital beamformer design is completely omitted in the following part. Consequently, the variables to be determined are $\left\lbrace \boldsymbol{f}_k \right\rbrace ^K_{k=1}$. 

Since the perfect CSI is unavailable, we cannot figure out the achievable rate in \eqref{Rk}. However, from \eqref{PerfectCSIProblem}, we see that the sum-rate can be maximized indirectly by maximizing the signal-to-interference-plus-noise ratio (SINR) of each user. Note that AoD in \eqref{channelmodel} is unknown and cannot be acquired by beam sweeping.  After beam sweeping, the best $K$ codewords for the $ K $ users can form a set $ \left\lbrace \boldsymbol{w}_k \right\rbrace_{k=1}^{K} $, where the angle range of the AoD corresponding to the best codeword of the $k$th user is denoted as $ \Omega_k $, for $k=1,2,\ldots,K$. Since $\Omega_k$ is only an angular range and not the exact AoD in~\eqref{channelmodel}, we obtain an approximation of the SINR for each user by sampling $\Omega_k$. Without loss of generality, we take the $ q $th user as an example for $q=1,2,\dots,K$. We approximate its SINR as
\begin{equation}\label{SINRq}
	{\rm {SINR}}_q \approx \frac{\frac{1}{M}\sum_{m=1}^{M} \left|\boldsymbol{a}(\phi_{q,m})_{\mathcal{S}_q}^H\boldsymbol{f}_q\right|^2}
	{\frac{1}{M}\sum_{k=1,k \neq q}^{K} \sum_{m=1}^{M} \left|\boldsymbol{a}(\phi_{q,m})_{\mathcal{S}_k}^H\boldsymbol{f}_k\right|^2+\sigma^2},
\end{equation}
where $M$ is the number of total samples in the angle range of each user AoD, and $\phi_{q,m} $ denotes the $m$th sampling point in $ \Omega_q $ for $m=1,2,\ldots,M$. Note that $\mathcal{S}_q $ denotes a index set of the non-zero entries of $ [\boldsymbol{F}_{\rm RF}]_{:,q} $. In fact, $ \boldsymbol{a}(\phi_{q,m})_{\mathcal{S}_q} $ denotes a new vector composed of entries of $ \boldsymbol{a}(\phi_{q,m}) $ whose indices are in $ \mathcal{S}_q $. Specifically, we approximate the desired signal power of the $ q $th user and the power of the interference to the $ q $th user by $ \frac{1}{M}\sum_{m=1}^{M}\lvert\boldsymbol{a}(\phi_{q,m})_{\mathcal{S}_q}^H\boldsymbol{f}_q\rvert^2 $ and $ \frac{1}{M}\sum_{k=1,k \neq q}^{K} \sum_{m=1}^{M} \lvert\boldsymbol{a}(\phi_{q,m})_{\mathcal{S}_q}^H\boldsymbol{f}_k\rvert^2 $, respectively.

It is seen from \eqref{SINRq} that the beamforming vectors of all users are coupled, which is difficult to solve. Therefore, we decouple these beamforming vectors based on the signal-to-leakage-and-noise ratio (SLNR) criterion to mitigate the multiuser interference~\cite{SLNR}. For example, for the $q$th user, the desired signal power and the power leaked from the $q$th user to all the other users can be approximated as $ \frac{1}{M}\sum_{m=1}^{M}\lvert\boldsymbol{a}(\phi_{q,m})_{\mathcal{S}_q}^H\boldsymbol{f}_q\rvert^2 $ and $\frac{1}{M} \sum_{k=1,k \neq q}^{K} \sum_{m=1}^{M} \lvert\boldsymbol{a}(\phi_{k,m})_{\mathcal{S}_q}^H\boldsymbol{f}_q\rvert^2 $, respectively. Then the optimization problem of $\boldsymbol{f}_q$ can be expressed as
  
\begin{subequations}\label{BeamNullingFrac}
	\begin{align}
		& \max_{ \boldsymbol{f}_{q} } \frac{\frac{1}{M} \sum_{m=1}^{M}\left|\boldsymbol{a}(\phi_{q,m})_{\mathcal{S}_q}^H\boldsymbol{f}_q\right|^2}
		{\frac{1}{M}\sum_{k=1,k \neq q}^{K} \sum_{m=1}^{M}  \left|\boldsymbol{a}(\phi_{k,m})_{\mathcal{S}_q}^H\boldsymbol{f}_q\right|^2 + \sigma^2} \label{ObjSLNR} \\
		&\ {\rm s.t.} \ \lvert \left[\boldsymbol{f}_{q}\right]_{n} \rvert  = \frac{1}{\sqrt {N_{\rm s}}}, n=1,2,\ldots,N_{\rm s}.	
	\end{align}
\end{subequations}
In fact, we deal with the total leaking power from the $q$th user to all the other users in \eqref{BeamNullingFrac} based on SLNR criterion, instead of dealing with the leaking power of all the other users to the $q$th user in~\eqref{SINRq}, which in turn leads to the suppression of multiuser interference and can increase the sum-rate in~\eqref{PerfectCSIProblem}.

To maximize~\eqref{ObjSLNR}, we should make the power of $ \boldsymbol{f}_q $ radiated in $ \Omega_q $ as large as possible and the power of $ \boldsymbol{f}_q $ radiated in $ \Omega_k(k \neq q) $ as small as possible. In fact, we can make $ \boldsymbol{f}_q $ form nulling in $ \Omega_k(k\neq q) $ with the minimum mainlobe offset to maximize the desired signal power in $ \Omega_q $. The analog-only beamformer design based on imperfect CSI is then transformed into an optimization problem of beam nulling design. However, the non-convex constraint on $ \boldsymbol{f}_q $ and the fractional form of \eqref{ObjSLNR} make \eqref{BeamNullingFrac} difficult to solve. To simplify~\eqref{BeamNullingFrac}, we introduce a weighting factor $ \lambda $ and rewrite it as
\begin{subequations}\label{BeamNullingfq}
	\begin{align}
		& \min_{\boldsymbol{f}_{q} } -\sum_{m=1}^{M}\left|\boldsymbol{a}(\phi_{q,m})_{\mathcal{S}_q}^H\boldsymbol{f}_q\right|^2
		+ \lambda\sum_{k=1,k \neq q}^{K} \sum_{m=1}^{M} \left|\boldsymbol{a}(\phi_{k,m})_{\mathcal{S}_q}^H\boldsymbol{f}_q\right|^2 \label{objective value}\\
		&\ {\rm s.t.}\ \ \big| \left[\boldsymbol{f}_{q}\right]_{n} \big|   = \frac{1}{\sqrt {N_{\rm s}}}, n = 1, 2,\ldots, N_{\rm s}. \label{BeamNullingConstraint}
	\end{align}
\end{subequations}
where the term $ \lambda M \sigma^2 $ is omitted as it is independent of $ \boldsymbol{f}_q $.
%This optimization can be solved via Manifold Toolbox, 
%named AMM (Accelerated Majorization-Minimization) 
%The manopt manifold toolbox can be used to solve the optimization problem in \eqref{BeamNullingfq}, but it takes a long time to converge. To speed up the convergence,
%we propose an iterative algorithm based on the MM framework.

\begin{algorithm}[!t]
	\caption{A-MM Beamforming Scheme Based on Imperfect CSI}
	\label{alg2}
	\begin{algorithmic}[1]
		\STATE \textbf{Input:} $ \left\lbrace \boldsymbol{w}_k \right\rbrace ^K_{k=1} $, $ \left\lbrace {\Omega}_k \right\rbrace ^K_{k=1} $, $\lambda$, $M$, $ N_{\rm BS} $, $ N_{\rm s} $.
		%\STATE Obtain $ \left\lbrace \left\lbrace \phi_{k,m}\right\rbrace^{M}_{m=1} \right\rbrace^{K}_{k=1} $ via sampling.
		\FOR{$q=1:K$}
		\STATE Initialize $ \boldsymbol{f}^{(0)}_{q} = \boldsymbol{w}_q$, $ i = 0 $.
		\REPEAT
		\STATE Obtain $ \boldsymbol{\Sigma}^{(i)}_{1} $ via~\eqref{sigma1}.
		\STATE Obtain $ \boldsymbol{\Sigma}^{(i)}_{2} $ via~\eqref{sigma2}.
		\STATE Obtain $ \boldsymbol{f}^{(i+1)}_{q}$ via~\eqref{coloseform}.
		\STATE $ i = i + 1 $.
		\UNTIL \textit{stop condition}  is satisfied.
		\STATE $ \boldsymbol{f}_q = \boldsymbol{f}^{(i)}_{q} $.
		\ENDFOR
		\STATE $ \boldsymbol{F}_{\rm RF} = {\rm blkdiag}\left\lbrace  \boldsymbol{f}_1, \boldsymbol{f}_2, \ldots, \boldsymbol{f}_K \right\rbrace  $.
		\STATE \textbf{Output:} $\boldsymbol{F}_{\rm RF}$.
	\end{algorithmic}
\end{algorithm}

To deal with the non-convex constant modulus constraint of \eqref{BeamNullingConstraint}, we resort to the  majorization–minimization (MM) method. The principle of the MM method to solve~\eqref{BeamNullingfq} is to find a majorized function of \eqref{objective value} that is easy to solve. Specifically, each element of the first term in \eqref{objective value} satisfies
\begin{equation}\label{term1}
	\begin{aligned}
		&\ \ \ -\left|\boldsymbol{a}(\phi_{q,m})_{\mathcal{S}_q}^H\boldsymbol{f}_q\right|^2 =-\boldsymbol{f}_q^H\boldsymbol{a}(\phi_{q,m})_{\mathcal{S}_q}\boldsymbol{a}(\phi_{q,m})_{\mathcal{S}_q}^{H}\boldsymbol{f}_q\\
		&\leqslant -(\boldsymbol{f}^{(i)}_{q})^H\boldsymbol{a}(\phi_{q,m})_{\mathcal{S}_q}\boldsymbol{a}(\phi_{q,m})_{\mathcal{S}_q}^{H}\boldsymbol{f}^{(i)}_{q}\\
		&\ \ \ -2\left( \boldsymbol{a}(\phi_{q,m})_{\mathcal{S}_q}\boldsymbol{a}(\phi_{q,m})_{\mathcal{S}_q}^{H}\boldsymbol{f}^{(i)}_{q}\right)^{H}\left( 
		\boldsymbol{f}_q-\boldsymbol{f}^{(i)}_{q}\right)\\
		&=-2\left( \boldsymbol{a}(\phi_{q,m})_{\mathcal{S}_q}\boldsymbol{a}(\phi_{q,m})_{\mathcal{S}_q}^{H}\boldsymbol{f}^{(i)}_{q}\right)^{H}\boldsymbol{f}_q + c_1,
	\end{aligned}
\end{equation}
where $ \boldsymbol{f}^{(i)}_{q} $ is the solution obtained in the $ i $th iteration of MM algorithm, and $ c_1 $ is a constant independent of $ \boldsymbol{f}_q $. For the second term in~\eqref{objective value}, combining the modulus constraint of $ \boldsymbol{f}_q $ and Lemma~1 proposed in~\cite{song2015optimization}, each element of the second term satisfies
\begin{equation}\label{term2}
	\begin{aligned}
		&\ \ \ \left|\boldsymbol{a}(\phi_{k,m})_{\mathcal{S}_q}^H\boldsymbol{f}_q\right|^2 =\boldsymbol{f}_q^H\boldsymbol{a}(\phi_{k,m})_{\mathcal{S}_q}\boldsymbol{a}(\phi_{k,m})_{\mathcal{S}_q}^{H}\boldsymbol{f}_q\\
		&\leqslant \boldsymbol{f}_{q}^H\boldsymbol{f}_{q} + (\boldsymbol{f}^{(i)}_{q})^H
		\left(\boldsymbol{I}_{N_{\rm s}} - \boldsymbol{a}(\phi_{k,m})_{\mathcal{S}_q}\boldsymbol{a}(\phi_{k,m})_{\mathcal{S}_q}^{H}\right)\boldsymbol{f}^{(i)}_{q}\\
		&\ \ \ + 2{\rm Re}\left\lbrace \boldsymbol{f}_{q}^H \left(\boldsymbol{a}(\phi_{k,m})_{\mathcal{S}_q}\boldsymbol{a}(\phi_{k,m})_{\mathcal{S}_q}^{H}-\boldsymbol{I}_{N_{\rm s}}  \right) \boldsymbol{f}^{(i)}_{q}\right\rbrace\\
		&= 2{\rm Re}\left\lbrace \boldsymbol{f}_{q}^H \left(\boldsymbol{a}(\phi_{k,m})_{\mathcal{S}_q}\boldsymbol{a}(\phi_{k,m})_{\mathcal{S}_q}^{H}-\boldsymbol{I}_{N_{\rm s}}  \right) \boldsymbol{f}^{(i)}_{q}\right\rbrace +c_2,
	\end{aligned}
\end{equation}
where $ c_2 $ is another constant independent of $ \boldsymbol{f}_q $.
\begin{figure}[t]
\centering
\includegraphics[width=3 in]{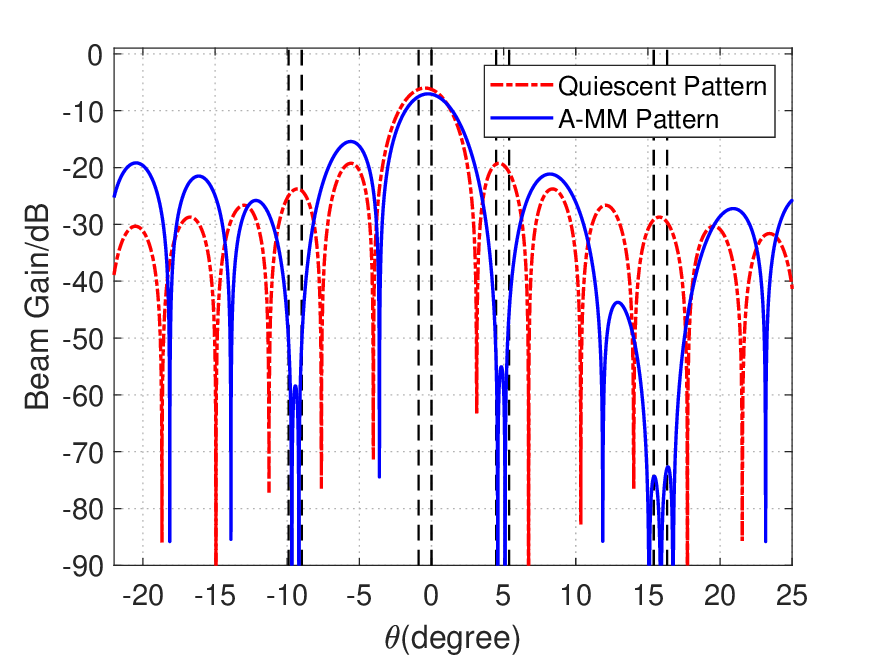}
\caption{Beam pattern comparison of A-MM and quiescent beam.}
\label{fig:Beampattern}
\end{figure}

% dangling modifier
%Leaving out the constant terms
After ignoring the constant terms, the majorized problem of \eqref{BeamNullingfq} can be expressed as
\begin{subequations}\label{BeamNullingfqEq}
	\begin{align}
		& \min_{\boldsymbol{f}_{q} } {\rm Re}\left\lbrace \boldsymbol{f}_{q}^{H} \left( 
		-\boldsymbol{\Sigma}^{(i)}_{1}
		+\lambda\boldsymbol{\Sigma}^{(i)}_{2}\right)  \right\rbrace  \\
		&\ {\rm s.t.}\ \lvert \left[\boldsymbol{f}_{q}\right]_{n} \rvert  = \frac{1}{\sqrt {N_{\rm s}}}, n =1,2,\ldots, N_{\rm s},
	\end{align}
\end{subequations}
where
\begin{subequations}
	\begin{align}
		\boldsymbol{\Sigma}^{(i)}_{1} &= \sum_{m=1}^{M}\boldsymbol{a}(\phi_{q,m})_{\mathcal{S}_q}\boldsymbol{a}(\phi_{q,m})_{\mathcal{S}_q}^{H}\boldsymbol{f}^{(i)}_{q},\label{sigma1}\\
		\boldsymbol{\Sigma}^{(i)}_{2} &= \sum_{k \neq q}^{K}\sum_{m=1}^{M}
		\left(\boldsymbol{a}(\phi_{k,m})_{\mathcal{S}_q}\boldsymbol{a}(\phi_{k,m})_{\mathcal{S}_q}^{H}-\boldsymbol{I}_{N_{\rm s}}  \right) \boldsymbol{f}^{(i)}_{q}.\label{sigma2}
	\end{align}
\end{subequations}
We can derive the closed-form solution to \eqref{BeamNullingfqEq} as
\begin{equation}\label{coloseform}
	\left[\boldsymbol{f}^{(i+1)}_{q}\right]_{n} = e^{j\angle\left\lbrace\left[\boldsymbol{\Sigma}^{(i)}_{1}-\lambda\boldsymbol{\Sigma}^{(i)}_{2}\right]_{n} \right\rbrace }, n =1,2,\ldots, N_{\rm s}.
\end{equation}
Then the $ (i+1) $th update of $ \boldsymbol{f}_{q} $ is completed by the MM method. The overall analog-only beamforming scheme based on imperfect CSI is summarized in \textbf {Algorithm \ref{alg2}}, which is named as the analog-only MM (A-MM) scheme.

%%%%%%%%%%%%%%%%%%%%%%%%%%%%%%%%%%%%%%%%%%%%%%%%%%%%%%%%%%%%%%%%%%%%%%%%%%%%%%%%%%%%%%%%%%%%%%%%%%%%%%%%%%%%%%%%%%%%%%%%%%%%%%%%%%%%%%%%%%%%%%%%%%%%%%%%%%%%%%%%%%%

\section{Simulation Results}\label{sec.SimulationResults}
First we compare the beam pattern of the A-MM scheme and the quiescent beam pattern that is essentially the initial beam pattern without nulling. We assume that the BS employs a half-wavelength spaced ULA with $ N_{\rm BS} = 128 $ antennas and $ N_{\rm RF}=4 $ RF chains to communicate with $ K = 4 $ single antenna users, where the number of antennas in each subarray is set as $ N_{\rm s} = 32 $. The angle range of each user is set as $ \Omega_1 = [ -0.9^{\circ}, 0^{\circ}]  $, $ \Omega_2 = [4.5^{\circ},  5.4^{\circ}] $, $ \Omega_3 = [-9.9^{\circ}, -9^{\circ}] $ and $ \Omega_4 = [15.4^{\circ}, 16.3^{\circ}] $, respectively. The weighting factor is set to be $ \lambda = 1000 $. Fig.~\ref{fig:Beampattern} shows the comparison of beam patterns for the first user. It is seen that A-MM forms nulls in desired angle range while preserving the gain in $ \Omega_1 $, where the nulling depths reach $ -55.1 $dB, $ -58.5 $dB and $ -72.7 $dB in $ \Omega_2 $, $ \Omega_3 $ and $ \Omega_4 $, respectively. Therefore, the multiuser interference is effectively mitigated.

Then we compare the proposed P-WMMSE scheme, the A-MM scheme, fully digital beamforming (Fully Digital), HySBD~\cite{wu2018hybrid} and TSH~\cite{alkhateeb2015limited} in terms of sum-rate. We extend HySBD and TSH to the partially-connected structure and further extend TSH to a noisy scenario with imperfect CSI for fair comparisons with the A-MM scheme.

\begin{figure}[t]
\centering
\includegraphics[width=3 in]{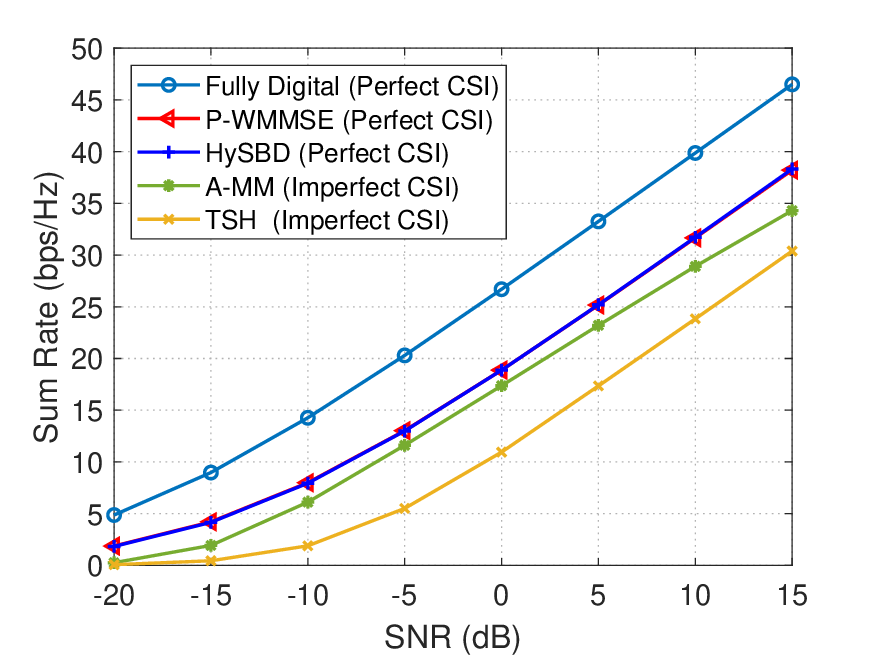}
\caption{Sum-rate performance v.s. SNR for different schemes.}
\label{fig:Achievable Sum Rate SNR}
\end{figure}

Fig.~\ref{fig:Achievable Sum Rate SNR} shows the sum-rate achieved by the five schemes with different SNR. We assume that the BS employs a half-wavelength spaced ULA with $ N_{\rm BS} = 512 $ antennas and $ N_{\rm RF}=4 $ RF chains. The channel between the BS and each user is generated according to~\eqref{channelmodel}, where the number of channel paths is set up to $ L_k = 3 $ with one line-of-sight (LoS) path and two non-line-of-sight (NLoS) paths. The complex gain of the LoS path is assumed to obey $ \alpha_{k,1} \sim \mathcal{CN}(0, 1) $ and that of the NLoS paths obeys $ \alpha_{k,l} \sim \mathcal{CN}(0, 0.01)$. We set $ \lambda = 1000 $ and perform Monte Carlo simulations based on 2000 random channel implementations. As shown in Fig.~\ref{fig:Achievable Sum Rate SNR}, based on the perfect CSI, the proposed P-WMMSE can achieve the same performance as HySBD, while avoiding SVD for high-dimensional channel matrix $ \boldsymbol{H} \triangleq[\boldsymbol{h}_1,\cdots,\boldsymbol{h}_K ]^H \in \mathbb{C}^{K \times N_{\rm BS}} $ and therefore reducing the computational complexity. The performance of A-MM reaches $ 92.0\% $ and $ 91.1\% $ of that of HySBD with perfect CSI at SNR = 5dB and SNR = 10dB, respectively. Besides of the better performance of A-MM than that of TSH, A-MM does not need to acquire the CSI of effective channels and therefore reduces the pilot overhead and omits the channel estimation therein.

% Fig.~\ref{fig:Achievable Sum Rate K} shows the sum-rate achieved by the five schemes varying with $ K $. We set $ N_{\rm BS} = 512 $, SNR = 10dB and $ \lambda = 1000 $. The number of antennas in each subarray is $ N_{\rm s} = \lfloor \frac{N_{\rm BS}}{N_{\rm RF}} \rfloor$ and other parameter settings are the same as those of Fig.~\ref{fig:Achievable Sum Rate SNR}. Similar to Fig.~\ref{fig:Achievable Sum Rate SNR}, P-WMMSE can achieve almost the same performance as HySBD, and the proposed A-MM outperforms TSH.
\begin{figure}[t]
\centering
\includegraphics[width=3 in]{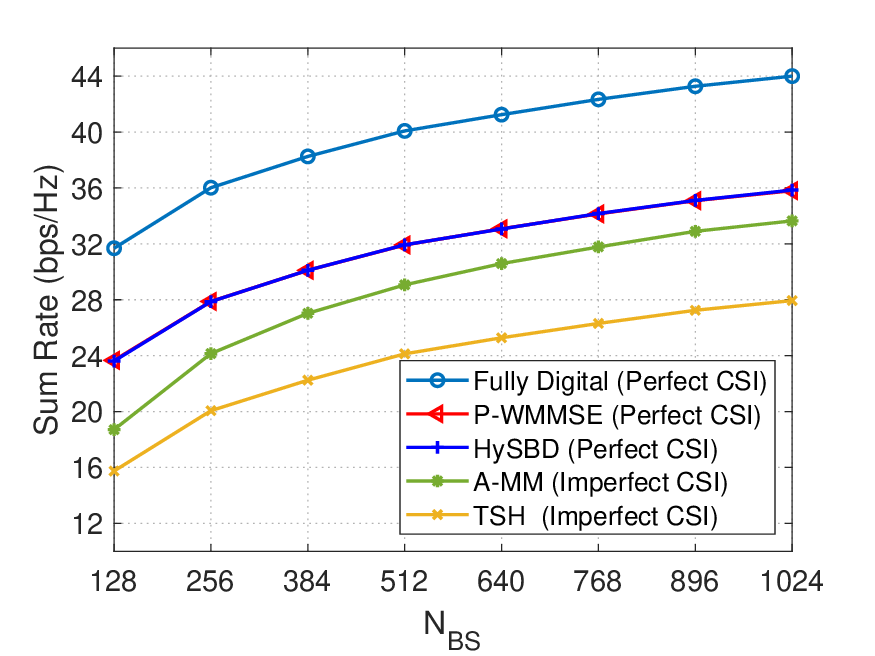}
\caption{Sum-rate performance v.s. $ N_{\rm BS} $ for different schemes.}
\label{fig:Achievable Sum Rate NBS}
\end{figure}

Fig.~\ref{fig:Achievable Sum Rate NBS} shows the sum-rate achieved by the five schemes with different $ N_{\rm BS}$. We set $ K = 4 $ and SNR = 10dB. The sum-rate of the five schemes grows with the increase of  $ N_{\rm BS} $, which is mainly contributed by the gain harvested by large-scale antenna arrays. From the figure, the performance gap between A-MM and the hybrid beamforming schemes based on perfect CSI gets smaller as $ N_{\rm BS} $ increases. The reasons are given as follows. With the increase of $ N_{\rm BS} $, the array gain and the number of codewords in the codebook increase, which increases the beam resolution  and reduces the errors in beam sweeping. Once the width of the mainlobe and sidelobe of the beam becomes narrow, we can reduce the influence of nulling on the mainlobe and the peak beam gain, which improves the signal power received by the user. Therefore, the sum-rate performance can be effectively improved by increasing $ N_{\rm BS} $ if the hardware complexity is supported.

%%%%%%%%%%%%%%%%%%%%%%%%%%%%%%%%%%%%%%%%%%%%%%%%%%%%%%%%%%%%%%%%%%%%%%%%%%%%%%%%%%%%%%%%%%%%%%%%%%%%%%%%%%%%%%%%%%%%%%%%%%%%%%%%%%%%%%%%%%%%%%%%%%%%%%%%%%%%%%%%%%
\section{Conclusion}\label{sec.Conclusion}
%\section*{Acknowledgment}
In this paper, we have proposed a low-complexity hybrid beamforming scheme based on perfect CSI and an analog-only beamforming scheme based on imperfect CSI for multiuser mmWave massive MIMO systems with partially-connected structure. Future work will be continued with the focus on multiuser beamforming to improve the sum-rate performance for different structures.

\bibliographystyle{IEEEtran}
\bibliography{IEEEabrv,IEEEexample}

\end{document}